\documentclass{sig-alternate-05-2015}

\usepackage[utf8]{inputenc}

\usepackage[english]{babel}
\usepackage{color}
\usepackage{hyperref}
\usepackage{xspace} % needed for \madmom

\usepackage{tikz}

\usepackage{caption}
\usepackage{listings}

\def\papertitle{madmom: a new Python Audio and Music Signal Processing Library}
\title{\papertitle}

\def\paperauthorA{Sebastian Böck}
\def\paperauthorB{Filip Korzeniowski}
\def\paperauthorC{Jan Schlüter}
\def\paperauthorD{Florian Krebs}
\def\paperauthorE{Gerhard Widmer}

\hypersetup{
    pdftitle={\papertitle},
    pdfauthor={\paperauthorA, \paperauthorB, \paperauthorC, \paperauthorD, \paperauthorE},
    pagebackref=true,
    colorlinks=true,
    urlcolor=black,
    linkcolor=black,
    citecolor=black,
    breaklinks=true,
    hyperfootnotes=true,
    bookmarksnumbered,
    pdfstartview=XYZ
}
\numberofauthors{1} %  in this sample file, there are a *total*
\author{
\alignauthor
\hspace{-3mm}\paperauthorA \dag \titlenote{To whom correspondence should be addressed; Email: \email{\texttt{sebastian.boeck@jku.at}}},
\paperauthorB \dag, \paperauthorC \ddag, \paperauthorD \dag, \paperauthorE \dag \ddag\\
    \affaddr{\dag~Department of Computational Perception, Johannes Kepler University Linz, Austria}\\
    \affaddr{\ddag~Austrian Research Institute for Artificial Intelligence (OFAI), Vienna, Austria}\\
}
\newcommand{\madmom}{\emph{madmom}\xspace}

\newcommand{\python}{\emph{Python}\xspace}
\newcommand{\numpy}{\emph{NumPy}\xspace}
\newcommand{\ndarray}{\emph{ndarray}\xspace}
\newcommand{\scipy}{\emph{SciPy}\xspace}
\newcommand{\cython}{\emph{Cython}\xspace}
\newcommand{\Processor}{\emph{Processor}\xspace}
\newcommand{\Processors}{\emph{Processors}\xspace}
\newcommand{\process}{\emph{process}\xspace}
\newcommand{\OutputProcessor}{\emph{OutputProcessor}\xspace}
\newcommand{\SequentialProcessor}{\emph{SequentialProcessor}\xspace}
\newcommand{\ParallelProcessor}{\emph{ParallelProcessor}\xspace}

\newcommand{\Signal}{\emph{Signal}\xspace}
\newcommand{\FramedSignal}{\emph{FramedSignal}\xspace}

\newcommand{\STFT}{\emph{ShortTimeFourierTransform}\xspace}

\newcommand{\Spectrogram}{\emph{Spectrogram}\xspace}

\newcommand{\Filterbank}{\emph{Filterbank}\xspace}

\newcommand{\MFCC}{\emph{MFCC}\xspace}
\newcommand{\Chroma}{\emph{Chroma}\xspace}

\newcommand{\pickle}{\emph{pickle}\xspace}

\begin{document}

\maketitle

% ============================= ABSTRACT
\begin{abstract}
In this paper, we present \madmom, an open-source audio processing and music information retrieval (MIR) library written in Python. \madmom features a concise, \numpy-compatible, object oriented design with simple calling conventions and sensible default values for all parameters, which facilitates fast prototyping of MIR applications.
Prototypes can be seamlessly converted into callable processing pipelines through \madmom's concept of \Processors, callable objects
that run transparently on multiple cores.
Processors can also be serialised, saved, and re-run to allow results to be easily reproduced anywhere. 

Apart from low-level audio processing, \madmom puts emphasis on musically meaningful high-level features.
Many of these incorporate machine learning techniques and \madmom provides a module that implements some in MIR commonly used methods such as hidden Markov models and neural networks.
Additionally, \madmom comes with several state-of-the-art MIR algorithms for onset detection, beat, downbeat and meter tracking, tempo estimation, and piano transcription.
These can easily be incorporated into bigger MIR systems or run as stand-alone programs.
\end{abstract}



%  Use this command to print the description
%%% removed for arxiv <----------------------------------------
%\printccsdesc

% We no longer use \terms command

%%% removed for arxiv <----------------------------------------
%\keywords{Music Information Retrieval, Audio Analysis, Signal Processing, Open Source, Python}

%%% removed for arxiv <----------------------------------------
%\pagebreak
% ============================= INTRO
\section{Introduction}
\label{intro}

% MIR allgemein
Music information retrieval (MIR) has become an emerging research area over the last 15 years. Especially audio-based MIR has become more and more important, since the amount of available audio data in the last years exploded beyond being manageable manually. 

Most state-of-the-art audio-based MIR algorithms consist of two components: First, low-level features are extracted from the audio signal (\emph{feature extraction} stage), and then the features are analysed (\emph{feature analysis} stage) to retrieve the requested information. Most current MIR systems incorporate machine learning algorithms in the feature analysis stage, with neural networks currently being the most popular and successful ones \cite{Schlueter2013, Korzeniowski2014, Boeck2014, Boeck2015}.

Numerous software libraries have been proposed over the years to facilitate research and development of applications in MIR. Some libraries concentrate on low-level feature extraction from audio signals, such as \emph{Marsyas} \cite{Tzanetakis2000}, \emph{YAAFE} \cite{Mathieu2010} and \emph{openSMILE} \cite{Eyben2013}. Others also include higher level feature extraction such as onset and beat detection as for example in the \emph{MIRtoolbox} \cite{Lartillot2007}, \emph{Essentia} \cite{Bogdanov2013} and \emph{librosa} \cite{McFee2015}. However, to our knowledge, there exist no library that also includes machine learning components (except \emph{Marsyas} \cite{Tzanetakis2000}, which contains two classifiers), although machine learning components are crucial in current MIR applications.

Therefore, we propose \madmom, a library that incorporates low-level feature extraction \emph{and} high-level feature analysis based on machine learning methods. This allows the construction of the full processing chain within a single software framework, making it possible to build standalone programs without any dependency on other machine learning frameworks. Moreover, \madmom comes with several state-of-the-art systems including their trained models, for example for onset detection \cite{Schlueter2013, Eyben2010}, tempo estimation \cite{Boeck2015}, beat estimation \cite{Boeck2014, Korzeniowski2014}, downbeat estimation \cite{Boeck2016}, and piano transcription.

\madmom is written in Python, which has become the language of choice for scientific computing for many people due to its free availability and its simplicity to use. The code is released under BSD license and pre-trained models are released under the CC BY-NC-SA 4.0 license.

% Previously, most MIR algorithms used a combination of different tools in different programming languages, often in low-level languages such as C++, incorporating dedicated feature extraction tools such as \emph{YAAFE} \cite{Mathieu2010}, openSMILE \cite{Eyben2013}, or built in frameworks such as \emph{Marsyas}\cite{Tzanetakis2000} requiring a deep understanding of signal processing workflows.

% When development of \madmom started, the only tools available in high-level languages were built around the Matlab programming environment, e.g.\ the \emph{MIRtoolbox}\cite{Lartillot2007}, but none in Python\footnote{Technically this is not correct, since \emph{librosa} is exactly 10 days older.}, which has become the language of choice for scientific computing for many people.
% In the meantime, \emph{Essentia}\cite{Bogdanov2013} (offering Python bindings) and \emph{librosa}\cite{McFee2015} were introduced.
% Both have a lot in common with \madmom, but we will outline \madmom's different
% focus below.

%\pagebreak
% =================== Design Goals
\subsection{Design and Functionality}

% % ======== intuitive use
% \subsubsection{Ease of use}
% 
% We want \madmom to be easy to use, thus we built everything around what most Python users are familiar with: \numpy's multi-dimensional \ndarray\cite{Walt2011}.
% \numpy arrays provide an intuitive way of handling multi-dimensional data and offer efficient methods for numerical operations and array manipulation.

% ======== OOP
\subsubsection{Object-oriented programming}
\label{sec:OOP}

\madmom follows an object-oriented programming (OOP) approach.
We encapsulate everything in objects which are often designed as subclasses of \numpy's \ndarray, offering all array handling routines inherited from \numpy\cite{Walt2011} with additional functionality.
This compactly bundles data and meta-data (e.g.,\ a \Spectrogram
and its \emph{frame rate}) and simplifies meta-data handling for the user. 
% Furthermore, objects can be instantiated from different class types using the same input parameter.we can use the same input parameter of a constructor for different class
% Furthermore, this allows for automatic resolution of the instantiation chain of classes, meaning classes can be instantiated from multiple inputs.

% ======== 
\subsubsection{Rapid prototyping}

\madmom aims at minimizing the turnaround time from a research idea to a software prototype. To this means, object instantiation is made as simple as possible: E.g., a \emph{log Mel-spectrogram} object can be instantiated with one line of code by only providing the path to an audio file. \madmom automatically creates all the objects in between using sensible default values.

% ======== 
\subsubsection{Simple conversion into runnable programs}

Once an audio processing algorithm is prototyped, the complete workflow should be easily transformed into a run\-nable standalone program with a consistent calling interface. This is implemented using \madmom's concept of \Processors. For details we refer to Section \ref{sec:library_desc}.

% \madmom provides \Processor wrappers around the respective \emph{data classes}, making the transformation often as easy as substituting all \emph{data classes} with the respective \Processors and removing the first positional argument.

% ======== 
\subsubsection{Machine learning integration}

We aim at a seamless integration of machine learning methods without the need of any third party modules. We limit ourselves to testing capabilities (applying pre-trained models), since it is impossible to keep up with newly emerging training methods in the various machine learning domains. Models that have been trained in an external library should be easily be converted to an internal \madmom model format.

% ======== 
\subsubsection{State-of-the-art features}

Many existing libraries provide a huge variety of low-level features but few musically meaningful high-level features. \madmom tries to close this gap by offering high-quality state-of-the-art feature extractors for downbeats, beats, onsets, tempo, piano transcription, etc..

% ======== Reproducible Research
\subsubsection{Reproducible research}

In order to foster reproducible research, we want to be able to save and load the specific settings used to obtain the results for a certain experiment. In \madmom this is implemented using Python's own \emph{pickle} functionality which allows to save an entire processing chain (including all settings) to a file.

% ======== dependencies
\subsubsection{Few dependencies}

\madmom is built on top of three excellent and wide-spread libraries:
\numpy\cite{Walt2011} provides all the array handling subroutines for \madmom's \emph{data classes}.
\scipy\cite{Jones2001} provides optimised routines for the fast Fourier transform (FFT), linear algebra operations and sparse matrix representations. Finally, \cython\cite{Behnel2011} is used to speed up time critical parts of the library by automatically generating C code from a Python-like syntax and then compiling and linking it into extensions which can be transparently used from within Python.
These libraries are the only installation and runtime dependencies of \madmom besides the \python standard library itself, supported in version 2.7 as well as 3.3 and newer.

% ======== Multi-Core
%%% removed for arxiv <----------------------------------------
%\pagebreak
\subsubsection{Multi-core capability}

We designed \madmom to be able to exploit the multi-core capabilities of modern computer architectures, by providing function to run several programs or \Processors in parallel.

% ======== Documentation
\subsubsection{Extensive documentation}

All source code files contain thorough documentation following the \numpy format.
The complete API reference, instruction on how to build and install the library, as well as interactive \emph{IPython} \cite{Perez2007} notebooks can be found online at {\small\tt\href{http://madmom.readthedocs.io}{http://madmom.readthedocs.io}}.
The documentation is build automatically with \emph{Sphinx}\footnote{\small\tt\href{http://www.sphinx-doc.org}{http://www.sphinx-doc.org}}.

% ======== Development
\subsubsection{Open development process}

We follow an open development process and the source code and documentation of our project is publicly available on GitHub: {\small\tt\href{http://github.com/CPJKU/madmom}{http://github.com/CPJKU/madmom}}.
%The source code is released under the liberal BSD license.
To maintain high code quality, we use continuous integration testing via TravisCI\footnote{\small\tt\href{http://www.travis-ci.org}{http://www.travis-ci.org}}, code quality tests via QuantifiedCode\footnote{\small\tt\href{http://www.quantifiedcode.com}{http://www.quantifiedcode.com}}, and test coverage via Coveralls\footnote{\small\tt\href{http://www.coveralls.io}{http://www.coveralls.io}}.

% ============================= LIBRARY DESCRIPTION
\section{Library Description}
\label{sec:library_desc}
In this section, we will describe the overall architecture of \madmom, its packages as well as the provided standalone programs.

\madmom's main API is composed of classes, but much of the functionality is
implemented as functions (in turn used internally by the classes). This way,
\madmom offers the `best of both worlds': concise interfaces exposed through
classes, and detailed access to functionality through functions.
In general, the classes can be split in two different types: the
so called \emph{data classes} and \emph{processor classes}.

\textbf{Data classes} represent data entities such as audio signals or spectrograms. They are implemented as subclasses of \numpy's \ndarray, and thus offer all array handling routines inherited directly from \numpy (e.g.\ transposing or saving the data to file in either binary or human readable format). These classes are enriched by additional attributes and expose additional functionality via methods. 

% E.g.\ the \Signal class has a \emph{sample\_rate} attribute, thus there is no need to manually carry this information in an extra variable and pass it to functions which need it.
% Furthermore, \emph{data classes} try to automatically instantiate an object of the expected class if their input parameter is not of the correct class. E.g.\ a \Spectrogram can be instantiated either from a \Signal or a simple file name.

\textbf{Processor classes} exclusively store information on how to process data, i.e.,\ how to transform one data class into another (e.g.,\ from an (audio-)\Signal into a \Spectrogram). In order to build chains of transformations, each data class has its corresponding processor class, which implements this transformation. This enables a simple and fast conversion of algorithm prototypes to callable processing pipelines.
% Since each data class can be created by a transformation of a lower-level data class (e.g.,\ from a \Signal into a \Spectrogram), each data class has a corresponding processor class which implements this transformation. This enables a simple and fast conversion of algorithm prototypes to callable processing pipelines.

% Like normal Python objects, processor classes can be \emph{pickled} to save and restore them later. This facilitates reproducible experiments: it is easy to store and load the exact processing pipeline (including parameters) that created the results. 
%For more technical details about \Processors, we refer to Section \ref{madmom.processors}.

% ============================= PACKAGES
\subsection{Packages}

The library is split into several packages, grouped by functionality.
For a detailed description including examples of usage please refer to the library's documentation.
% For simpler accessibility, important and often used classes and their corresponding \Processors are imported directly into the library's main namespace.

% ======== Processors
\subsubsection{madmom.processors}
\label{madmom.processors}

\Processors are one of the fundamental building blocks of \madmom. Each \Processor accepts a number of processing parameters and must provide a \process method, which takes the data to be processed as its only argument and defines the processing functionality of the \Processor. An \OutputProcessor extends this scheme by accepting a second argument which defines the output and can thus be used to write the output of an algorithm to a file. All \Processors are callable, making it easy to use them interchangeably with normal functions. Further, the \Processor class provides methods for saving and loading any \Processor to a file -- including all parameters -- using Python's own \pickle library. This facilitates the reproducibility of an experiment.

Multiple \Processors can be combined into a processing chain using a \SequentialProcessor or \ParallelProcessor, which either execute the chain sequentially or in parallel, using multiple CPU cores if available.
% \madmom automatically processes them sequentially or in parallel, using multiple CPU cores if possible. It is easy to define a complete (sequential) processing chain by instantiating a \SequentialProcessor with a list of (predefined) \Processors.

% ======== audio
\subsubsection{madmom.audio}

The \emph{madmom.audio} package includes basic audio signal processing and ``low-level'' functionality.
% It comprises everything necessary to built algorithms (e.g.\ signal transformations) that cannot be evaluated directly (e.g.\ positions of onsets). For this package, we just describe the \emph{data classes}. As mentioned earlier, each \emph{data class} has a corresponding \Processor. These are not described separately, since they expose the same functionality.

% Signal & FramedSignal
The \Signal and \FramedSignal classes are used to load an audio signal and chop it into (overlapping) frames.
Following \madmom's automatic instantiation approach, both classes can be instantiated from any object up the instantiation hierarchy -- including a simple file name. \madmom supports almost any existing audio and video formats -- provided \emph{ffmpeg}\footnote{\small\tt\href{http://www.ffmpeg.org}{http://www.ffmpeg.org}} is installed -- and transparently converts sample rates and number of channels if needed.

\Signal is a subclass of \ndarray with additional attributes like \emph{sample rate} or \emph{number of channels}. 
% This way it bundles the signal data with its meta-data.
\FramedSignal supports float hop sizes, making it possible to build systems with an arbitrary frame rate -- independently of the signal's sample rate -- and ensures that all frames are temporally aligned, even if computed with different frame sizes.

% STFT & Spectrogram
The \STFT and \Spectrogram classes represent the complex valued STFT and magnitudes respectively.
They are the key classes for spectral audio analysis and provide windowing, automatic circular shifting (for correct phase) and zero-padding. Both are independent of the data type (integer or float) of the underlying \Signal, resulting in spectrograms of the same value range. A \Spectrogram can be filtered with a \Filterbank (e.g.\ Mel, Bark, logarithmic), which in turn can be parametrised to reduce the dimensionality or transform the spectrogram into a logarithmically spaced pitch representation closely following the auditory model of the human ear. \madmom also provides standard \MFCC and \Chroma features.
% \TODO{cite Emilia?}.

% \TODO{neuerdings mehr und mehr direkt auf Audiospektrogrammen gelernt, statt 1001 hand-crafted Features; trotzdem gibt es so grundlegende Dinge wie MFCCs, Chroma, etc.}
 
% ======== Features
\subsubsection{madmom.features}

The \emph{madmom.features} package includes ``high-level'' functionality which are related to certain MIR tasks, such as beat tracking or onset detection.
% Features and descriptors provided here can usually be evaluated directly (e.g.\ beat positions).
\madmom's focus is to provide musically meaningful and descriptive features rather than a vast number of low to mid-level features. At the time of writing this paper, \madmom contains state-of-the-art features for onset detection, beat and downbeat tracking, rhythm pattern analysis, tempo estimation and piano transcription.

All features are implemented as \Processors without a corresponding \emph{data class}. Users can thus use the provided functionality and build algorithms on top of these features. For most of the features, \madmom also provides stand-alone programs with a consistent calling interface to process audio files (see Section \ref{sec:programs}).

% ======== Evaluation
\subsubsection{madmom.evaluation}

All features come with code for evaluation. The implemented metrics are those commonly found in the literature of the respective field.

%%% removed for arxiv <----------------------------------------
%\pagebreak
% ======== Machine Learning Stuff
\subsubsection{madmom.ml}
\label{madmom.ml}

%In recent years, machine learning techniques applied for all kinds of MIR tasks.

Most of today's top performing music analysis algorithms incorporate machine learning, with neural networks being the most universal and successful ones at the moment.
\madmom includes Python implementations of commonly used machine learning techniques, namely \emph{Gaussian Mixture Models} (GMM), \emph{Hidden Markov Models} (HMM), and different types of \emph{neural networks} (NN), including feed forward, recurrent, and convolutional layers, various activation functions and special purpose units such as \emph{long short-term memory} (LSTM).

\madmom provides functionality to use these techniques without any dependencies on third-party modules, but does not contain training algorithms.
This decision was made on purpose since the library's main focus is on applying machine learning techniques to MIR, rather than providing an extensive set of learning techniques.
However, trained models can be easily converted to be compatible with \madmom, since neural network layers usually are simply defined as a set of weights, biases and an activation function they apply to the input data.
%Some functionality is borrowed from other projects (e.g.\ GMMs from the well-known \emph{scikit-learn} \cite{Pedregosa2011} package), whereas other parts are written from scratch.
%\madmom includes a very efficient HMM implementation done in Cython.
%Fitting of GMMs still requires the \emph{scikit-learn} package and is called transparently, but when saving the fitted GMMs this dependency is left out. This allows users to use the full power of \emph{scikit-learn} for fitting GMMs without requiring it for evaluating them.

% ======== Models
\subsubsection{madmom.models}
\label{madmom.models}

\madmom comes with a set of pre-trained models which are distributed under a Creative Commons attribution non-commercial share-alike license, i.e.\ they can be freely used for research purposes as long as derivative works are distributed under the same license.
\madmom uses the exact same mechanism to save and load the models it uses for \Processors to be pickled.
% with all parameters to be re-run with the same setting.

% ======== Programs
\subsection{Standalone Programs}
\label{sec:programs}

\madmom comes with a set of standalone programs, covering many areas of MIR.
These programs are simple wrappers around the functionality provided by the \emph{madmom.features} package, and provide a simple and easy to use command line interface.
They are implemented as \Processors and can operate either in \emph{single} or \emph{batch} mode, processing single or multiple input files, respectively.
Additionally all programs can be \emph{pickled}, serialising all parameters in a way that the program can be executed later on with the exact same settings.

Table \ref{tab:programs} lists selected programs included in the library with the performance achieved at the annual \emph{Music Information Retrieval Evaluation eXchange} (MIREX)\footnote{\small\tt\href{http://www.music-ir.org/mirex/wiki/}{http://www.music-ir.org/mirex/wiki/}}, where MIR algorithms are compared on hidden test datasets.
We aggregated the results of all years (2006-2015), i.e., a rank 1 means that the algorithm is the best performing one of all submissions from 2006 until present. The outstanding results in Table \ref{tab:programs} highlight the state-of-the-art features \madmom provides.
% In cases of better performing systems, it should be noted that these algorithms are not publicly available under an open source license.

\begin{table}[h!tp]
\caption[]{\textbf{Ranks of the programs included in \madmom for the MIREX evaluations, results aggregated over all years (2006-2015). Asterisks indicate pending submissions. 
%%% removed for arxiv <----------------------------------------
%\protect\footnotemark
}}
\label{tab:programs}
\begin{center}
\begin{tabular}{llcc}
Program & Task & Year & Rank \\
\hline
% onsets
\hline
CNNOnsetDetector \cite{Schlueter2013} & onset & 2013 & 1 \\
OnsetDetector \cite{Eyben2010} & onset & 2013 & 2 \\
%OnsetDetectorLL \cite{Boeck2012a} & onset & 2012 & 4 \\
%ComplexFlux \cite{Boeck2013a} & onset & 2012 & 5 \\
%SuperFlux \cite{Boeck2013} & onset & 2012 & 6 \\
% beats
\hline
BeatTracker \cite{Boeck2011} & beat MCK & 2015 & 1 \\
DBNBeatTracker \cite{Boeck2014} & beat SMC & 2015 & 1 \\
CRFBeatDetector \cite{Korzeniowski2014} & beat MAZ & 2015 & 1 \\
% downbeats
\hline
GMMPatternTracker \cite{Krebs2013} & downbeat & 2015 & 2 \\
DBNDownBeatTracker \cite{Boeck2016} & downbeat & 2016 & * \\
% tempo
\hline
TempoDetector \cite{Boeck2015} & tempo & 2015 & 1 \\
% chord
\hline
CRFChordTranscriptor & chord & 2016 & * \\
% transcription
\hline
PianoTranscriptor & transcription & 2016 & * \\
\end{tabular}
\end{center}
\label{default}
\end{table}%
%
%%% removed for arxiv <----------------------------------------
%\footnotetext{Note to reviewers: work on integrating existing state-of-the-art music segmentation and chord transcription algorithms based on convolutional neural nets (CNNs) has started and is expected to be finished in time for the camera-ready submission. Additionally, the results of the pending submissions in Table 1 will be available as well.}

% ============================= CONCLUSION
%%% removed for arxiv <----------------------------------------
%\pagebreak
\section{Conclusion}

This paper gave a short introduction into \madmom, its design principles and library structure.
%\madmom provides unique features such as \Processors that can be used to build 
Up-to-date information on functionality can be found in the project's online documentation at {\small\tt\href{https://madmom.readthedocs.io}{https://madmom.readthedocs.io}} and source code repository at {\small\tt\href{https://github.com/CPJKU/madmom}{https://github.com/CPJKU/madmom}}. 

Future work aims at including a streaming mode, i.e.\ providing online real-time processing of audio signals in a memory efficient way instead of processing whole audio files at a time.
In addition, we will gradually extend the set of features and algorithms, as well as add tools to automatically convert models that have been trained with popular machine learning libraries such as Lasagne \cite{lasagne}.

% ============================= ACKS
\section{Acknowledgments}
This work is supported by the European Union Seventh Framework Programme FP7 / 2007-2013 through the \linebreak GiantSteps project (grant agreement no.~610591) and the Austrian Science Fund (FWF) project Z159.

\bibliographystyle{abbrv}
%{\small
\bibliography{acmmm-madmom}
%}

% That's all folks!
\end{document}